\documentstyle[12pt,fleqn]{article}
\input epsf
\epsfverbosetrue
\begin{document}
\newcommand{\beq}{\begin{equation}}
\newcommand{\eeq}{  \end{equation}}
\newcommand{\bea}{\begin{eqnarray}}
\newcommand{\eea}{  \end{eqnarray}}
\newcommand{\bit}{\begin{itemize}}
\newcommand{\eit}{  \end{itemize}}
\newcommand{\beps} {\mbox{\boldmath $\epsilon$}}
\newcommand{\sbeps}{\mbox{\scriptsize \boldmath $\epsilon$}}

\title{Boundary contributions to the semiclassical traces \\ of 
the baker's map\thanks{Submitted to Nonlinearity}}

\author{F. Toscano,$^1$ 
        R. O. Vallejos$^{1,2}$\thanks{Permanent address: 
Centro Brasileiro de Pesquisas F\'{\i}sicas, Rua Xavier Sigaud 150, 
CEP 22290-180, RJ, Rio de Janeiro, Brazil.}
        and 
        M. Saraceno$^{3}$\thanks{Permanent address: 
Departamento de F\'{\i}sica, Comisi\'on Nacional de Energ\'{\i}a 
At\'omica (CNEA), Av. del Libertador 8250, 1429 Buenos Aires,
Argentina.}}
        
\maketitle

\noindent
\centerline{\it $^1$Centro Brasileiro de Pesquisas F\'{\i}sicas}
\centerline{\it Rua Xavier Sigaud 150, CEP 22290-180, RJ, Rio de
                Janeiro, Brazil}
\centerline{\it $^2$Service de Physique Th\'eorique,} 
\centerline{\it Commissariat \`a l'Energie Atomique--Centre d'Etudes
                Nucl\'eaires de Saclay,}
\centerline{\it F--91191 Gif--sur--Yvette CEDEX, France}
\centerline{\it $^3$Centre Emile Borel, Institut Henri Poincar\'e,}
\centerline{\it 11 Rue Pierre et Marie Curie, 75231 Paris Cedex,
France}

\vspace{1cm}

\centerline {\bf Abstract}

We evaluate the leading asymptotic contributions to the traces
of the quantum baker's map propagator. Besides the usual Gutzwiller
periodic orbit contribution, we identify boundary paths giving 
rise to anomalous $\log \hbar$ terms. 
Some examples of these anomalous terms are calculated both
numerically and analytically.

\newpage

\section{Introduction}
\par

The usual derivation of the periodic orbit trace formula for chaotic
systems extracts as stationary phases the actions of isolated periodic
orbits while the amplitudes are given by the contributions of their
hyperbolic neighbourhoods. However, when the orbits lie on
discontinuities or at the boundary of classically allowed regions,
anomalous contributions are expected.\cite{NUS}
We analize and calculate the origin of these contributions for the
simplest piecewise--linear map:  the baker's transformation.

In spite of many generic properties and its inherent simplicity, the
baker's map has shown some "anomalies", both in the statistical
properties of the quasi--energy spectrum\cite{OCT} and in the
asymptotic behaviour of the lowest traces of the propagator.\cite{SAV}
It is now understood that these anomalies are related to the
discontinuous way in which the map transforms phase space and to the
singular nature of the fixed  point:\cite{LAK} it lies on the
discontinuity and does not have a whole hyperbolic vicinity (in this
sense, all the other periodic points are regular). In Ref.~\cite{SAV}
it was shown that, when the map is quantized, and its semiclassical
traces considered in the context of periodic orbit theory, the
contributions coming from singular symbols (related to the fixed point
and its repetitions) show anomalous terms with a $\log \hbar$
dependence. Some indications of anomalous behaviour for regular symbols
(related to regular periodic orbits) were also communicated.

Our main result is that not only the singular symbols contribute
anomalously to the trace but {\em every} regular symbol also does. We
trace these contributions to classical paths (not periodic orbits)
whose action is stationary at the vertices of classically allowed
(hypercube--shaped) domains.

The paper is organized as follows.  In Section 2 we briefly review the
classical and quantum features of the map that will be relevant for
further discussions.  Section 3 describes the manipulations on the
iterated propagator that lead to an exact symbolic decomposition of its
traces. The exact contribution to the trace from each symbol sequence
results in a path summation formula which, following a procedure
suggested in Ref.~\cite{LUO1}, is converted by Poisson transformation
into a multiple infinite number of (continuous variables) path
integrals in phase space. This is the starting point for the
approximations to be made.

In Section 4 we analize the different stationary paths that contribute
to these integrals and their charachteristic $\hbar$ dependence, thus
obtaining their asymptotic behaviour.  We arrive at the result that,
for each symbol, the asymptotic behaviour is ruled by more than one
path. The periodic orbit (labeled with that symbol) contributes with
the usual Gutzwiller term and a {\em vertex} path contributes with a
logarithmic term.  Some examples of these anomalous terms are
calculated, both analytically and numerically.  Further constant terms
arising from other stationary paths are identified but are very
hard to calculate. Section 5 contains the concluding remarks.

\section{Review of the classical and quantum baker's map}

In this section we display the well known classical and quantum
ingredients of the baker's that will be needed for further analysis.

The classical baker's transformation is an area preserving,
piecewise--linear map of the unit square (periodic boundary conditions
are assumed) defined as
\beq
p_{1}=\frac{1}{2}(p_{0}+\epsilon_0) ~,~~~ 
q_{1}= 2q_{0}-\epsilon_0 ~;
\label{bakclas}
\eeq
where $\epsilon_{0}=[2q_{0}]$, the integer part of $2q_{0}$.  This map
is known to be uniformly hyperbolic, the stability exponent for orbits
of period $L$ being $L\log 2$.  Moreover it admits a useful description
in terms of a complete symbolic dynamics. A one to one correspondence
between phase space coordinates and  binary sequences,
\beq
(p,q) \leftrightarrow 
\ldots 
\epsilon_{-2} 
\epsilon_{-1} 
\cdot 
\epsilon_{0} 
\epsilon_{1} 
\epsilon_{2} 
\ldots
~~~,~ \epsilon_i=0,1 ~,
\label{symbol}
\eeq
can be constructed in such a way that the action of the map is
conjugated to a shift map.  The symbols are assigned as follows:
$\epsilon_i$ is set to zero (one) when the $i$--th iteration of 
$(p,q)$ falls to the left (right) of the line $q=1/2$, i.e. 
$[2q_i]=\epsilon_i$. 
Reciprocally, given an itinerary
$\ldots 
\epsilon_{-2} 
\epsilon_{-1} 
\cdot 
\epsilon_{0} 
\epsilon_{1} 
\epsilon_{2} 
\ldots$, the related phase point is obtained through the especially
simple binary expansions
\beq
q=\sum_{i=0}^{\infty} \frac{\epsilon_i}{2^{i+1}} ~,~~~
p=\sum_{i=1}^{\infty} \frac{\epsilon_{-i}}{2^i} ~.
\label{pqsym}
\eeq
Once the dynamics has been mapped to a shift on binary sequences it is
very easy to analize the dynamical features of the map.  In particular,
periodic points are associated  to infinite repetitions of {\em finite}
sequences of symbols. It will be convenient for later purposes to
introduce a vector notation for these sequences so that $\beps=
(\epsilon_{0},\epsilon_{1},\ldots,\epsilon_{L-1})^t$ (the superscript
means transposition).
We also denote the positions and momenta of a periodic trajectory 
of length $L$ as
${\bf q}^*$=$(q^*_0,q^*_1,\ldots,q^*_{L-1})^t$ and 
${\bf p}^*$=$(p^*_0,p^*_1,\ldots,p^*_{L-1})^t$. 
For a given $\beps$, the initial point on the trajectory is obtained 
by considering 
a periodic itinerary of length $L$ in (\ref{pqsym})\cite{SAR} 
\beq
q_0^{*}=\frac{1}{2^{L}-1}
        \sum_{i=0}^{L-1} { 2^{L-1-i} \, \epsilon_i}  ~,~~~
p_0^{*}=\frac{1}{2^{L}-1}
        \sum_{i=0}^{L-1} { 2^{ i}    \, \epsilon_i}  ~.
\eeq
The other points are obtained by shifting cyclically the binaries
$\epsilon_i$ in the expression above. In a compact form,
\beq
{\bf q}^*= {\bf A}^{-1} \, \beps  ~,~~~
{\bf p}^*= ({\bf A}^t)^{-1} \, {\bf S} \, \beps ~.
\label{perpoint}
\eeq
The matrix ${\bf A}^{-1}$ is directly related to the matrix 
${\bf S}$ of a cyclic shift, 
${\bf S} \cdot  (a_0,a_1,\ldots a_{L-1})^t = 
              (a_1,\ldots ,a_{L-1},a_0)^t$,
\beq
{\bf A}^{-1}= \frac{1}{2^{L}-1}
              \sum_{i=0}^{L-1}{ 2^{L-1-i} \, {\bf S}^i} ~,
\label{matrixAinv}
\eeq
and embodies all the symmetries of the periodic trajectories.\cite{SAR}
Its inverse will play an important role in the quantum and
semiclassical analysis and has the simple form,
\beq
{\bf A} ={\bf 2} - {\bf S} ~.
\label{matrixA}
\eeq
Due to its piecewise linear nature, the baker's map admits a (mixed)
generating function which is a piecewise bilinear form,
\beq
W_{\epsilon_{0}}(p_{1},q_{0})=
2p_{1}q_{0} - \epsilon_{0}p_{1} -\epsilon_{0}q_{0} 
~,~~~ \epsilon_{0}=0,1 ~.
\label{genf}
\eeq
It is not defined on the whole space $p_1$--$q_0$ but on the classically
allowed domains
\beq
R_0=[0,1/2]\times[0,1/2] ~~~ \mbox{and} ~~~ R_1=[1/2,1]\times[1/2,1] ~.
\label{domains}
\eeq 
Once the generating function is defined, actions can be assigned
to the periodic orbits.\cite{OZS} In our notation they read
\beq
S_{\sbeps}= \beps^t       \, {\bf A}^{-1} \, \beps 
            = ({\bf S}{\bf p}^*)^t \, {\bf A}      \, {\bf q}^*  ~.
\label{action}
\eeq
\vspace{1pc}

With respect to the quantum map, we will follow the original
quantization of Balazs and Voros\cite{BAV}, as later modified in Ref.
\cite{SAR} to preserve in the quantum map all the symmetries of its
classical counterpart. In the mixed representation the baker's
propagator can be written as an $N \times N$ block--matrix ($N$ even):
\beq
\langle p_{m}|B_N|q_{n} \rangle= 
                \left( \begin{array}{cc}
                           G_{N/2} &      0            \\
                              0    & G_{N/2}
                       \end{array}
                \right) ~,
\label{qbdef}
\eeq
where position and momentum eigenvalues run on a discrete mesh with
step $1/N=h$ ($h$ = Planck's constant), so that
\beq
q_{n}=(n+1/2)/N ~,~~~ p_{m}=(m+1/2)/N ~,~~~ 0 \le n,m \le N-1 ~;
\label{grid}
\eeq
and $G_N$ is the antiperiodic Fourier matrix, which transforms 
from the $q$ to the $p$ basis,
\beq
G_{N}=\langle p_{m}|q_{n} \rangle=(1/\sqrt N)e^{-2\pi i N p_m q_n} ~.
\label{Fourier}
\eeq
This propagator\cite{OZS}  
has the standard structure of quantized linear symplectic maps,
\beq
\langle p_{m}|B|q_{n} \rangle =
\left\{ 
\begin{array}{cl}
\sqrt{2/N} \, e^{-i 2\pi N W_{0}(p_{m},q_{n})} &  
             \mbox{if} ~~ (p_{m},q_{n})\, \in \, R_0 \\
\sqrt{2/N} \, e^{-i 2\pi N W_{1}(p_{m},q_{n})} &  
             \mbox{if} ~~ (p_{m},q_{n})\, \in \, R_1 \\
0                                             &  
                                     \mbox{otherwise} ~.
\end{array}
\right.
\label{VanVleck}       
\eeq
In this quantization, only those transitions are
allowed that respect the rule $[2p_m]=[2q_n]$, a reflection of the 
classical shift property.

\section{Path integral formulation}

The iteration of the propagator in the mixed representation is simply
the expansion of a finite matrix product, which we write as
\beq
\langle p_{L}|B^{L}|q_0\rangle=
\sum
\langle p_{L}|B|q_{L-1}\rangle
\langle q_{L-1}|p_{L-1}\rangle \cdots 
\langle p_{2}|B|q_{1}\rangle 
\langle q_{1}|p_{1}\rangle
\langle p_{1}|B|q_0\rangle    ~,
\eeq     
where the summation is over repeated  phase space coordinates.  To
simplify notation we have dropped the subindices labeling the discrete
values of the coordinates in (\ref{grid}) in favor of a subindex
labeling the (discrete) time on the trajectory.  When (\ref{VanVleck})
is taken into account the trace of the iterated propagator can be
written as a sum over paths of length $L$
\beq
\mbox{Tr}B^{L}=\sum_{\gamma} A_{\gamma} \, e^{i W_{\gamma}/\hbar} ~,
\eeq
where $A_{\gamma}$ is the amplitude and $W_{\gamma}$ the action of a 
path $\gamma=(q_0,p_1,q_1,p_2,\cdots,q_{L-1},p_L)$ 
which satisfies the rule $[2q_i]=[2p_{i+1}]$. By grouping the paths
according to their symbolic itinerary
$\beps=(\epsilon_0,\epsilon_1,\ldots,\epsilon_{L-1})$, the trace can be
further decomposed as
$\mbox{Tr}B^{L}= \sum_{\sbeps} \mbox{Tr}B_{\sbeps}^{L}$, with
\beq
\mbox{Tr}B_{\sbeps}^{L}= 
\frac{2^{L/2}}{N^L}
\sum 
e^{-2\pi iN \, [
W_{\epsilon_{0}}(q_0,p_{1})-q_{1}p_{1}+
W_{\epsilon_{1}}(q_{1},p_{2})-\cdots - q_{L-1}p_{L-1}+ 
W_{\epsilon_{L-1}}(q_{L-1},p_{L})-q_0p_{L}]}   ~.
\label{pathnu}
\eeq
This decomposition breaks up the trace of the propagator into partial
sums each one labeled by a symbolic code. Contrary to the usual {\sl
semiclassical} procedure, this decomposition is now exact. Each symbol,
instead of corresponding semiclassically to a Gutzwiller contribution
from a periodic orbit is given by a sum over paths that share a common
symbolic code. The asymptotic evaluation of these sums can now be
attempted focusing on one symbol at a time. Thus from now on we
restrict the analysis to the partial trace $\mbox{Tr}B_{\sbeps}^{L}$,
i.e. we take $\beps$ fixed.
 
As a function of {\em continuous} coordinates,
the quadratic phase in (\ref{pathnu}) has the property of being 
stationary over the {\em only} periodic trajectory carrying the
label $\beps$ (for $N$ big enough, a discrete path arbitrarily
close to this trajectory can be found, so the stationary phase 
picture also holds in the discrete case). 
A first attempt at the semiclassical evaluation of
(\ref{pathnu})\cite{OZS} consisted in replacing the sums by
integrals which can then be evaluated in the stationary phase
approximation, neglecting boundary contributions. 
(Special care has to be
taken of the anomalous contributions of the fixed point and its
repetitions: these trajectories lie on a vertex of the domain of
integration, where the approximations stated before are not
applicable.)
In the case of regular
orbits, i.e.  
$\beps \ne (0,0,\cdots,0)$ and $\beps \ne (1,1,\cdots,1)$,
this procedure leads to the usual semiclassical expression
\beq
\mbox{Tr}B_{\sbeps}^{L}\approx
\frac{2^{L/2}}{2^{L}-1} \,  e^{2\pi i N S_{\sbeps}} ~,
\eeq
where $S_{\sbeps}$ is the action of the (stationary phase) periodic
orbit given in (\ref{action}). This result, combined with an {\em ad
hoc} evaluation of the anomalous fixed points contributions led to a
fairly good reconstruction of the smoothed properties of the
spectrum.\cite{OZS} However, a more careful semiclassical analysis of
the first traces\cite{SAV} revealed the existence of anomalous
contributions, typically terms of the order $\log \hbar$. The presence
of these terms was then interpreted as coming from the singular orbits;
nevertheless, certain quantitative differences could not be explained,
moreover, anomalous behaviour of a {\em regular} orbit was also
suggested.\cite{SAV}

In this paper we explain the reason for the failure of previous
attempts to obtain asymptotic expressions for the traces which is
intimately linked to the discreteness of the coordinate grid.  Another
path--not a trajectory--exists that renders the phase stationary. Small
discrete displacements about this path produce a (leading order)
variation of the phase which is an integer multiple of $2\pi$. This is
a path of {\em quasi}--stationary phase, related to {\em aliasing\/};
it lies at a vertex of the space of paths, and thus escapes the usual
stationary phase treatment.  This phenomenon occurs for {\em all}
symbols and therefore all traces are anomalous in this sense.  The
correct trace expansion for each symbol should then be written as
\beq
\mbox{Tr}B_{\sbeps}^{L}\approx
\frac{2^{L/2}}{2^{L}-1} \, e^{2\pi i N S_{\sbeps}} +
          a_{\sbeps} \log N + b_{\sbeps} + \ldots ~,
\label{expansion}
\eeq
where the first term will be absent for 
$\beps={\bf 0}$ or ${\bf 1}$ ($a_{\sbeps}$ and $b_{\sbeps}$ are 
complex constants). 

In order to clearly display the abovementioned features we 
follow Ref.~\cite{LUO1}
and transform the path summation into integrations via the Poisson
formula.  Making use of the identity
\beq
\sum_{i=0 \, (\frac{N}{2})}^{\frac{N}{2}-1 \, (N-1)} F(q_i)=
N  \sum_{K=-\infty}^{\infty} (-1)^K
\int_{0 \, (\frac{1}{2})}^{\frac{1}{2} \, (1)} \, 
e^{2\pi i N Kq} \, F(q) \, dq      ~,
\eeq
where the parenthesis indicate alternative ranges for the summation
(integration), we rewrite (\ref{pathnu}) as
\bea
\mbox{Tr}B_{\sbeps}^{L} & = & 2^{L/2}N^L 
\sum_{M_0} \cdots \sum_{M_{L-1}} 
\sum_{K_1} \cdots \sum_{K_{L}}
(-)^{M_0 + \ldots + M_{L-1} + K_1 + \ldots +K_L}  \times \nonumber \\
               & \times  &
\int_{R_{\epsilon_0}}      dq_0     dp_1 
\int_{R_{\epsilon_1}}      dq_1     dp_2 \cdots 
\int_{R_{\epsilon_{L-1}}}  dq_{L-1} dp_L \, 
e^{-2\pi iN\phi} ~.
\label{trazanu}
\eea
$M_{i},K_{i}$ come from Poisson transforming the variables $q_i,p_i$ 
and take all integer values. The phase $\phi$ is 
\bea
\phi & = &
W_{\epsilon_0}(q_0,p_1) - p_1 q_1 - M_0 q_0 - K_1 p_1  +
W_{\epsilon_1}(q_1,p_2) - p_2 q_2 - M_1 q_1 - K_2 p_2  + \cdots 
\nonumber \\
     &   &
\cdots +
W_{\epsilon_{L-1}}(q_{L-1},p_L) - p_L q_0 - M_{L-1}q_{L-1}-K_{L}p_L  ~.
\label{phi}
\eea
In a more compact form we write
\beq
\mbox{Tr}B_{\sbeps}^{L} =
\sum_{M,K} 
(-)^{M_0 + \ldots + M_{L-1} + K_1+ \ldots + K_L} 
I_{\epsilon,M,K} ~,
\label{Poisson}
\eeq
implicitly defining the integrals $I_{\epsilon,M,K}$.

This is the starting point of the semiclassical analysis 
(cf. (\ref{pathnu})). 
Standard stationary phase is now possible as
sums have been replaced by integrals--in an exact way.
The difficulty now is that we have to deal with an
infinite multiplicity of integrals and therefore it is essential
for further progress to identify among them those that 
contribute significantly to the asymptotic behaviour.
At this point our treatment diverges from that in 
Ref.~\cite{LUO1} 
where a truncation of the Poisson expansion was considered. On one 
side, the truncation misses constant terms which come from
{\em infinite} families of integrals (Appendix A). On the other, the
truncated set contains integrals which are not relevant
in the far asymptotic regime, as only a few specific ones are needed
(next section).

\section{The semiclassical trace formula}

In analizing the traces as given by (\ref{Poisson}) we follow the 
criterion 
that the most important contributions will come from those multiple
integrals having stationary phase paths lying in the interior of the
domains of integration (or on the boundary). Other cases 
will be oscillatory over the whole domain and thus contribute 
vanishingly.

For compactness we will use the matrix notation introduced
earlier.
The positions and momenta of a path of length $L$ will be 
respectively denoted
${\bf q}$=$(q_0,q_1,\ldots,q_{L-1})^t$ and 
${\bf p}$=$(p_1,\ldots,p_{L})^t$.
Analogously we define the vectors of 
integers coming from the Poisson transformation:
${\bf M}$=$(M_0,\ldots, M_{L-1})^t$ and 
${\bf K}$=$(K_1,\ldots, K_{L})^t$. For convenience of notation
notice that here ${\bf p}$ is shifted with respect to that
defined in Section 2. All vectors and matrices will be understood
to be $L$--dimensional.

The quadratic character of the generating function (\ref{genf}) can be
used to express the phase $\phi$ (\ref{phi}) of a path contributing to
the partial trace $\mbox{Tr} B_{\sbeps}^L$, and corresponding to
Poisson integers ${\bf M}$ and ${\bf K}$, as
\beq
\phi = {\bf p}^t \, {\bf A} \, {\bf q} - 
       {\bf p}^t \, ( \beps + {\bf K}) -  
       {\bf q}^t \, ( \beps + {\bf M})  ~;
\eeq
${\bf A}$ and its inverse were defined in 
(\ref{matrixAinv}, \ref{matrixA}). 
A stationary phase path $(\tilde{\bf q},\tilde{\bf p})$ 
is then a solution of the equations
$\nabla_{\bf q} \, \phi=0$, $\nabla_{\bf p} \, \phi=0$,
that is,
\beq
\tilde{\bf q}  = {\bf A}^{-1}     \, (\beps+ {\bf K}) ~,~~~
\tilde{\bf p}  = ({\bf A}^t)^{-1} \, (\beps+ {\bf M}) ~.
\label{nablazero}
\eeq
Equivalently, these stationary points could have been identified in the
discrete picture (\ref{pathnu}) by requiring the quasistationarity
of the action $W$, i.e. 
$\nabla_{\bf q} \, W= {\bf M}$, $\nabla_{\bf p} \, W={\bf K}$.
The advantage of having introduced the Poisson's transformation is
that it is now clear how to calculate the contribution of each stationary 
path.
For fixed $\beps$, (\ref{nablazero}) gives one solution for each choice 
of {\bf K} and {\bf M}.
The quadratic form, shifted to this stationary point becomes
\beq
\phi = ({\bf p}-\tilde{\bf p})^t \, {\bf A} \,
       ({\bf q}-\tilde{\bf q}) - 
       \tilde{\bf p}^t \, {\bf A} \, \tilde{\bf q} ~.       
\eeq
The point $(\tilde{\bf q},\tilde{\bf p})$ may fall inside, on the
vertices or faces, or outside the region of integration $R_{\epsilon_0}
\times R_{\epsilon_1} \times \ldots  \times R_{\epsilon_{L-1}}$.  There
are only three kinds of solutions (\ref{nablazero}) that do not fall
outside the domain of integration:
\bit
\item[(R)]
Interior paths. These paths coincide with the regular periodic 
trajectories (\ref{perpoint}) and are obtained by setting 
{\bf K} and {\bf M} to zero for $\beps \ne \bf{0},\bf{1}$. 
The corresponding coordinates are $\tilde{\bf q}={\bf q}^*$ 
and $\tilde{\bf p}={\bf S} \, {\bf p}^*$. 
The stationary value of the phase, $\tilde{\phi}\/$, is the 
classical action of the trajectory, i.e.  
$\tilde{\phi}=S_{\sbeps}\/$ (see Section 2).
\item[(V)]
Vertex paths. By choosing 
${\bf K}=({\bf A}  -1)\beps$ and
${\bf M}=({\bf A}^t-1)\beps$, 
we get the solution $\tilde{\bf q}=\beps$, $\tilde{\bf p}=\beps$. 
The corresponding value for the
phase, $\tilde{\phi}=\beps^t \, {\bf A} \, \beps $, is an
integer, and so
has no relevance in the phase in (\ref{trazanu}).
\item[(F)]
There are two {\em face} paths. Their coordinates are obtained
by mixing regular and vertex solutions, i.e. 
$(\tilde{\bf p},\tilde{\bf q})$=$(\beps ,{\bf q}^*)$ or
$({\bf p}^*,\beps)$.  
\eit
\noindent
It must be pointed out that in the case of the singular symbols 
$\beps = \bf{0},\bf{1}$ there is only one solution--the vertex one.
In Fig.~1 we show as an example the stationary paths for the case $L=3$ 
and the regular symbol $\beps = (0,0,1)$.

In the first case (R), the asymptotic evaluation
of the integral is straightforward. The stationary path coincides
with a regular periodic orbit and it is interior
to the integration domain, therefore boundary terms can be neglected.
The quadratic form can be brought to principal axes and evaluated
in a standard way. The result is simply 
\bea  
\mbox{I}_{\sbeps}^R & = &
2^{L/2} \, e^{2\pi i N S_{\sbeps}}  N^L \times
\int_{R_{\sbeps}} \, d{\bf p} \, d{\bf q} \,
e^{-2\pi i N \, ({\bf p} -\tilde{\bf p})^t \,
                {\bf A} \, ({\bf q}-\tilde{\bf q})} = \nonumber \\
                   & = &
\frac{2^{L/2}}{2^L-1} \, 
e^{2\pi i N S_{\sbeps}}  +  \mbox{terms decreasing with $N$} ~,
\label{gutz}
\eea
the usual Gutzwiller contribution 
(we have used $\det {\bf A}=2^L-1$).

For vertex (V) solutions this procedure is not possible because the 
neighbourhood
of the path (which only coincides with a periodic trajectory when
$\beps = \bf{0},\bf{1}$) is intersected by the boundaries of the 
hypercube
of integration. The integral to be evaluated can be brought to the
form (using the new coordinates ${\bf q} - \beps$ and 
${\bf p} - \beps$) 
\beq  
\mbox{I}_{\sbeps}^V =
2^{L/2} N^L \times
\int_{[0,\frac{1}{2}]^{2L}} \, d{\bf p} \, d{\bf q} \,
e^{-2\pi i N \, 
{\bf p}^t \, {\bf J}_{\sbeps} \, {\bf A} \, {\bf J}_{\sbeps} 
\, {\bf q}}  ~.
\label{intver}
\eeq
${\bf J}_{\sbeps}$ is a diagonal matrix with elements $\pm 1$,
$J_{ii}= 1- 2 \epsilon_{i-1} ~,~ 1 \le i \le L~,$
and contains the dependence on the symbol $\beps$.

(A comment about the invariant properties of (\ref{intver}) is in
order. The original path sum (\ref{pathnu}) is cyclically invariant
with respect to shifts on the symbol $\beps$ and also respects all the
symmetries of the map.\cite{SAV} Once the stationary paths have been
identified and their contributions evaluated it is important to verify
that these symmetries are maintained. In that case, only one
representative integral for each symbol and symmetry class needs
calculation. From the expression (\ref{intver}) it is possible to check
that this is indeed the case.)

To our knowledge, for $L > 2$, the leading asymptotic 
expression for the integral (\ref{intver}) cannot be evaluated
analytically. Appendix B presents the case $\mbox{I}_{(0,1)}^V$
as an example of the 
difficulties involved in such calculations. 
The analytical and numerical results for the shortest symbols 
indicate that $\mbox{I}_{\sbeps}^V$ displays the following
asymptotic behavior ($N\!\rightarrow\!\infty$),
\beq
I_{\sbeps}^V=
a_{\sbeps} \ln N + b_{\sbeps} + \mbox{decreasing terms}~,
\label{prediction}
\eeq
where $a_{\sbeps}$ and $b_{\sbeps}$ are (complex) constants.
For $L=1,2$ this result is obtained analytically, for $L=3$,
numerically.

The simplest example of a vertex integral is that associated to the
calculation of the first trace, worked out in the Appendix A.  In this
case the only stationary path is the vertex one, there are no interior
nor face paths. It is shown that asymptotically the vertex integrals
$I_{\sbeps=(0)}^V$ and $I_{\sbeps=(1)}^V$ give account of the $\log N$
term, but fail in reproducing the constant one which is present in
$\mbox{Tr} B_{\sbeps=(0)}$.  It is also proven that to obtain the
correct constant an additional infinite family of Poisson integrals
must be considered. These integrals, which provide a constant term,
contain continuous sets of stationary points {\em of the second
kind\/},\cite{BOW} i.e. their phases are stationary with respect to
displacements {\em along the faces} of the hypercube of integration. We
expect that in the general case ($L>1$) this behaviour will be
maintained: the vertex integral will provide the correct $\log N$ term,
but the constant $b_{\sbeps}$ will only be obtained when an infinite
family of Poisson integrals is calculated. So, even though constant
terms in the trace expansion are detected, its calculation is very hard
and will not be attempted here.

As to face solutions (F), there is numerical evidence that these
integrals will contribute with a constant.\cite{TOS} 
On the basis of the previous discussion, we subordinate
their study to further progress in the identification and
handling of the infinite families of integrals of the second kind that 
are also responsible
for the constant term.

In Figs. 2 and 3 we show, for the case $L=3$, numerical 
data concerning the asymptotic
behaviour of the exact traces 
as calculated by matrix multiplication 
(\ref{pathnu}) and the prediction
of our theory, i.e. the associated vertex integrals (\ref{prediction}).  
The representative traces $\mbox{Tr}B_{(0,0,0)}$ and
$\mbox{Tr}B_{(0,0,1)}$ are plotted as a function of even $N$ 
(in the second
case the Gutzwiller term has been substracted) together with 
some values of the vertex integrals $I_{(0,0,0)}^V$ and 
$I_{(0,0,1)}^V$ (much more time demanding than the
traces).
The straight lines are obtained by means of linear fits to
the numerical data (in the case of the traces only a subsequence of 
points with $N>2^{10}$ has been taken into account).
These figures show clearly that the vertex integral captures the
correct $\log N$ dependence but that the constant is incorrect. This is
exactly the same behaviour that is calculated analytically for the
cases $L=1,2$. The high quality of this agreement can be observed in
Table~\ref{tabla}, where a quantitative comparison between the
asymptotics of traces and integrals is displayed.
There we exhibit the coefficients $a_{\sbeps}$ of the
$\log N$ terms as extracted either from linear fits or
analytically (Appendices A and B, Ref.~\cite{SAV}).
A representative member of each symmetry class up to $L=3$ is shown.

Notice that the $\log N $ dependence does not dissappear in the full
traces and therefore it is not an artifice of the symbolic
decomposition. For instance for $L=3$ we have $\mbox{Tr}B^3= 2
\mbox{Tr}B_{(0,0,0)} + 6 \mbox{Tr}B_{(0,0,1)}$ and from Table~1 we get
$\mbox{Tr}B^3 \approx (-0.126-i \, 0.103) \log N$.

\section{Concluding remarks}

We have shown that stationary phase paths other than
periodic trajectories exist which significantly contribute 
to the semiclassical trace formula of the quantum baker's
map. 
The fact that these paths lie on vertices or faces of the allowed phase
space makes their contribution anomalous: they give rise 
to $\log \hbar$ and other terms. A theory is presented that not only
explains the origin of those anomalies but is capable of
quantitative predictions. In the case of the three
shortest traces, these predictions are compared
to the exact results showing excellent agreement. These
results encourage us to conjecture that {\em all} partial traces 
will display this behaviour. So, the semiclassical trace
formula for the baker's map becomes
\beq
\mbox{Tr}B^{L} = 
   \sum_{\sbeps}{ (a_{\sbeps} \log N +  b_{\sbeps}) }
 + \sum_{\sbeps \ne {\bf 0,1}}{ \frac{2^{L/2}}{2^{L}-1} 
                \, e^{2\pi i N S_{\sbeps}}    }
 + \mbox{decreasing terms} ~.
\eeq
Each coefficient $a_{\sbeps}$ involves the calculation of a
multidimensional integral, which, up to now, must be done numerically.
These integrals adopt the simple form of a quadratic phase which is
stationary on the vertex of an hypercube. This seems to be a
complicated task but maybe not unsolvable.  The calculation of the
constant terms $b_{\sbeps}$ in the expansion is even a more difficult
problem, as it involves the identification and handling of infinite
families of integrals (dominated by second kind stationary points).

Even though the results presented here have been obtained for a
particular quantization of the baker's map, we believe that these
results are extendible to any quantization (e.g. the {\em optical}
baker\cite{HKO}), as they are a consequence of the unusual nature of
the fixed point of the classical map, which lies on a discontinuity.
However, the particular values of the coefficients in the trace
formula may depend of the particular scheme used to quantize the
map.

The trace formula --without the $\log N$ terms-- was tested
numerically both for smoothed spectral properties using the density of
states\cite{OZS} and for individual eigenvalues using zeta-function
techniques.\cite{SAV} In both cases, for computational reasons, $N$ was
restricted to take relatively small values (e.g. $N=48$ in \cite{SAV},
$N$ up to 1024 in \cite{OZS}).
The agreement obtained was remarkable and can be explained by the fact
that for such low values the $\log N$ terms do not play a significant
role. It remains the open question, in view of our present results, if
that agreement is truly asymptotic or if it will deteriorate for larger
values of $N$.

\section*{Acknowledgements}

This work, which started at CNEA, greatly benefited from discussions at
the semester on ``Chaos and Quantization'' at the {\em Institut Henri
Poincar\'e}. In particular, discussions with A. Voros and A. M. Ozorio
de Almeida are gratefully acknowledged. We thank the following 
institutions for financial support: 
{\em Fundaci\'on} ANTORCHAS and CONICET (PID/3233/92); 
CLAF--CNPq (F.T., R.O.V.); CEA (R.O.V.); 
{\em Centre Emile Borel} (UMS 839/CNRS/UPMC) (M.S.).
Most part of the calculations were done on the CRAYJ90 at
NACAD--COPPE/UFRJ.


\appendix

\section*{Appendix A}

In this appendix we study the asymptotics of the first trace of 
the map. 
The result has been obtained in Ref.~\cite{SAV} with an 
{\em ad hoc} procedure to evaluate the sums. Here we show how to 
arrive at the same result but using stationary phase considerations
in the context of the Poisson 
expansion. This example 
illustrates the fact that the $\log \hbar$ term is
associated to the vertex integral but, in order to reproduce
the constant term an infinite family of Poisson integrals 
must be summed.

The starting point is the 
definition of $\mbox{Tr}B_{(0)}$, Eq.~(\ref{trazanu}),
\beq
\mbox{Tr}B_{(0)} =  
\sum_{M_0 =-\infty }^\infty 
\sum_{K_1 =-\infty }^\infty (-1)^{M_0 +K_1 }I_{(0),M_0 ,K_1 }  ~,
\label{apptrac}
\eeq
with
\bea
I_{(0),M_0,K_1} & = &
\sqrt{2}N
\int\limits_{0}^{1/2}dp_1
\int\limits_{0}^{1/2}dq_0 \, 
e^{-2 \pi iN (p_1 q_0 -M_0 q_0 -K_1 p_1)} \nonumber \\
            & =  &
\sqrt{2}N
\int\limits_{-M_0}^{-M_0 +1/2}dp_1
\int\limits_{-K_1}^{-K_1 +1/2}dq_0 \, e^{-2 \pi iN p_1 q_0} ~.
\label{appint}
\eea
A change of variables has been made to transfer 
the dependencies on
$M_0$ and $K_1$ to the domain of integration. Now the picture
is that of an hyperbolic phase, $e^{-2 \pi iN p_1 q_0}$, 
integrated over a domain that is an infinite union of (disjoint) 
squares of side $\frac{1}{2}$ 
(see Fig. 4).
We observe that, besides the stationary point in the
origin, all those points of the domain which lie on the 
axis $q_0$ or $p_1$ are stationary with respect to displacements
along those axes. These are non--isolated stationary points 
{\em of the second kind\/} and their contribution is non
negligible. So we restrict the integration to
those squares lying along the axes (shaded squares in Fig. 4), i.e.
\beq
\mbox{Tr}B_{(0)} \approx  I_{(0),0,0} +
\sum_{M = \pm 1 \pm 2 \ldots }(-1)^{M}I_{(0),M,0} +
\sum_{K = \pm 1 \pm 2 \ldots }(-1)^{K}I_{(0),0,K}  ~,
\eeq
or, by virtue of the $p$--$q$ symmetry $I_{(0),M,K}=I_{(0),K,M}$, 
\beq
\mbox{Tr}B_{(0)} \approx I_{(0),0,0} + 
2 \times \sum_{K=\pm 1,\pm 2,\ldots} (-1)^{K} I_{(0),0,K} ~.
\label{keyformula}
\eeq
This expression can be evaluated explicitly in the regime
$N \rightarrow \infty$. We begin by doing a first integration, 
over $q$, in (\ref{appint}): 
\bea
I_{(0),-M,-K} &=&
\frac{i}{\sqrt{2}\pi}\int_{M}^{M+1/2} \, 
\frac{dp}{p} \,
\{[\cos(2\pi N p(K+1/2))-1] -[\cos(2\pi N pK)-1]
     \nonumber \\ & &
-i [\sin(2\pi N p(K+1/2)) -\sin(2\pi N pK)]\} ~.
\eea
Each one of the integrals above
can be solved explicitly in terms of
the functions Ci$(z)$ and Si$(z)$,
$\int_{0}^{z} dp(\cos p-1)/p = \mbox{Ci}(z) -\gamma -\log z$, 
$\int_{0}^{z} dp \sin p /p   = \mbox{Si}(z)$ 
($\gamma$ is Euler's constant).\cite{ABS} 
A careful consideration
of the limits $M \rightarrow 0$, $N \rightarrow \infty$ leads
to
\bea
I_{(0),M=0,K=0} & \approx & 
-\frac{i\sqrt{2}}{2\pi}   \log N + \frac{\sqrt{2}}{4} 
-\frac{i\sqrt{2}}{2\pi} \left( \log \frac{\pi}{2} - \gamma \right) \\
I_{(0),M=0,-K} & \approx & 
\frac{\sqrt{2}}{2\pi}i \log \frac{K}{K-1/2} ~.
\eea
The contributions coming from each square can be added up 
by making use of the identity $\Pi_{n=1}^{\infty}(1-1/(2n)^2)=2/\pi$,
\cite{GRR} so that
\beq
\sum_{K=\pm 1,\pm 2,\ldots}(-1)^{K} \log \frac{K}{K-1/2}  = 
\log \frac{\pi}{4} ~.
\eeq
Replacing into (\ref{keyformula}) we obtain,
\beq
\mbox{Tr}B_{(0)} \approx 
-\frac{i\sqrt{2}}{2\pi}   \log N + \frac{\sqrt{2}}{4} +
 \frac{i\sqrt{2}}{2\pi} \left( \log \frac{\pi}{8} + \gamma \right) ~.
\eeq
Finally, as 
$\mbox{Tr}B_{(1)}=\mbox{Tr}B_{(0)}$, 
we have $\mbox{Tr}B=2\mbox{Tr}B_{(0)}$, the result in \cite{SAV}.

The difficulty in the computation of the constant term in general
is clearly demonstrated in this example.

\section*{Appendix B}
 
Here we calculate the asymptotic behavior of the integral
$\mbox{I}_{\nu=(1,0)}^{V}$ following the method developed
in Ref. \cite{SAV}. 
We start with the definition of this integral,
\begin{eqnarray*}
\mbox{I}_{(1,0)}^V&=&2 N^2\int_{[0,\frac{1}{2}]^4}
dp_1\,dp_2\,dq_0\,dq_1\;
e^{-2\pi iN[2p_1q_0+p_2q_0+2p_2q_1+p_1q_1]} \\
&=& 2\int_{[0,\frac{\sqrt{N}}{2}]^4}
dy_1\,dy_2\,dx_0\,dx_1\;
e^{-2\pi i[2y_1x_0+y_2x_0+2y_2x_1+y_1x_1]} ~.
\end{eqnarray*}
The last equality defines a smooth function of real $N$ whose 
derivative reduces to a triple integral over the four faces 
of the hypercube $[0,\sqrt{N}/2]^4$. As all four faces contribute 
equally by symmetry, we have
\begin{eqnarray}
\nonumber
\frac{d\mbox{I}_{(1,0)}^V}{dN}&=& 2\times\frac{1}{4\sqrt{N}}\times 4
\int_{[0,\frac{\sqrt{N}}{2}]^3}
dy_2\,dx_0\,dx_1\;
e^{-2\pi i[x_0(y_2+2\sqrt{N}/2)+x_1(2y_2+\sqrt{N}/2)]}\\
\nonumber
&=&2N\int_{[0,\frac{1}{2}]^3}
dp_2\,dq_0\,dq_1\;
e^{-2\pi iN[q_0(p_2+1)+q_1(2p_2+1/2)]} \\
\label{intdupla}
&=&2N\int_{0}^{1/2}\;dq_0
e^{-3\pi iNq_0/2}
\int_{0}^{1/2}\;dq_1
\frac{\sin\pi N(2q_1+q_0)/2}{\pi N(2q_1+q_0)}e^{-\pi i N(2q_1+q_0)}
\end{eqnarray}
(we have explicitly performed the $p_2$ integral). Making 
the coordinates change $q^{'}=q_1+q_0/2$ in the inner integral
and splitting its domain into two pieces we obtain
\begin{eqnarray}
\nonumber
& &\int_{0}^{1/2+q_0/2}\,dq^{'}\frac{\sin\pi Nq^{'}}{2\pi Nq^{'}}
e^{-2\pi iNq^{'}}-
\int_{0}^{q_0/2}\,dq^{'}\frac{\sin\pi Nq^{'}}{2\pi Nq^{'}}
e^{-2\pi iNq^{'}} \\
& &=\frac{1}{2\pi N}
\left[- \int_{0}^{1}\frac{dt}{t} \sin(\pi Ntq_0/2)e^{-2\pi iNtq_0/2}+
\int_{0}^{\pi N(1/2+q_0/2)}du\frac{\sin u}{u}e^{-2iu} \right]\;.
\end{eqnarray}
In the asymptotic limit ($N\rightarrow \infty$) the last integral gives
$\int_{0}^{\infty}e^{-2i\pi u}\sin/u \;du\;+{\cal O}(1/N)=
-i\ln(3)/2\;+{\cal O}(1/N)$. Replacing this result in the double integral 
(\ref{intdupla}), changing the order of integration and making the
$q_0$--integration explicitily, we arrive at
\begin{equation}
\frac{d\mbox{I}_{(1,0)}^V}{dN}=-\frac{1}{\pi^2N}\int_{0}^{1}
\left[\frac{e^{-i\pi N(t+3)q_0/2}}{t(t+3)}
-\frac{e^{-i\pi N3(t+1)q_0/2}}{3t(t+1)}-\frac{\log 3}{3} 
e^{-3\pi iNq_0/2}
 \right]_{q_0=0}^{q_0=1/2} ~.
\end{equation}
\noindent
The contribution from the $q_0=0$ endpoint to $d\mbox{I}_{(1,0)}^V/dN$
is
\begin{equation}
\frac{1}{\pi^2N}\int_{0}^{1}dt\left[ 
\frac{1}{t(t+3)}-\frac{1}{3t(t+1)}-\frac{\log 3}{3}\right]=
\frac{1}{N}\left(-\frac{\log 2}{3\pi^2}\right) ~.
\end{equation}
The contribution from the $q_0=1/2$ endpoint is of the same magnitude,
but oscillatory in $N$. Upon integration with respect to $N$ yields
\begin{equation}
\mbox{I}_{10}^V \approx -\frac{\log 2}{3\pi^2} \; \ln N \, + \,
\mbox{const.} 
\end{equation}
As in \cite{SAV} we have given up the determination of the additive
constant.

\newpage

\begin{figure} 
\epsfxsize=15.0cm
\epsfbox[55 169 482 339]{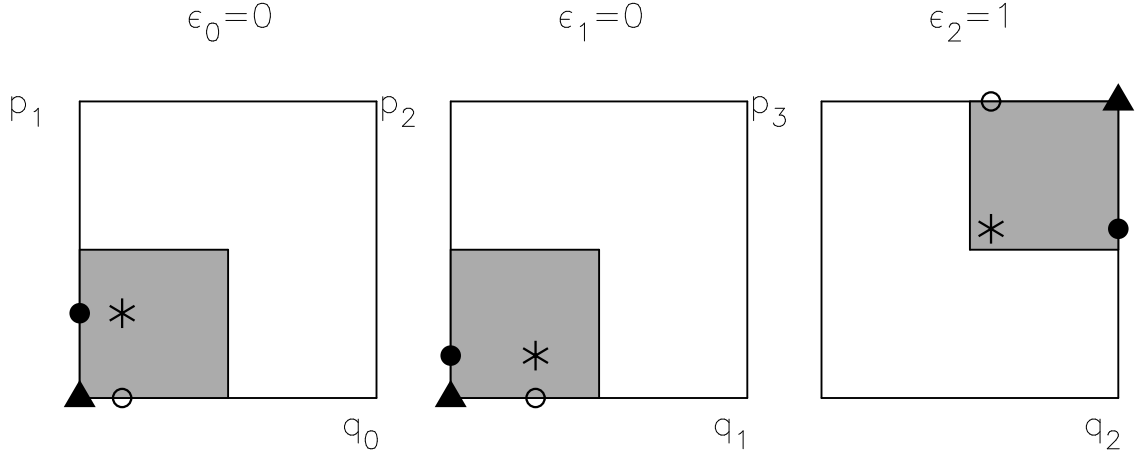}
\caption{Stationary phase paths associated to the
regular symbol $\beps =(0,0,1)$. The space of allowed paths is a
hypercube of side 1/2 (gray squares).  The interior path ($\ast$)
corresponds to a periodic trajectory and gives rise to the usual
Gutzwiller term in the trace formula. The vertex path ($\triangle$) and
its anomalous vicinity are responsible for the $\log \hbar$ term.
"Face" paths ($\circ$, $\bullet$) contribute with constant terms.}
\end{figure}

\newpage

\begin{figure} 
\epsfysize=15.0cm
\epsfbox[80 170 532 632]{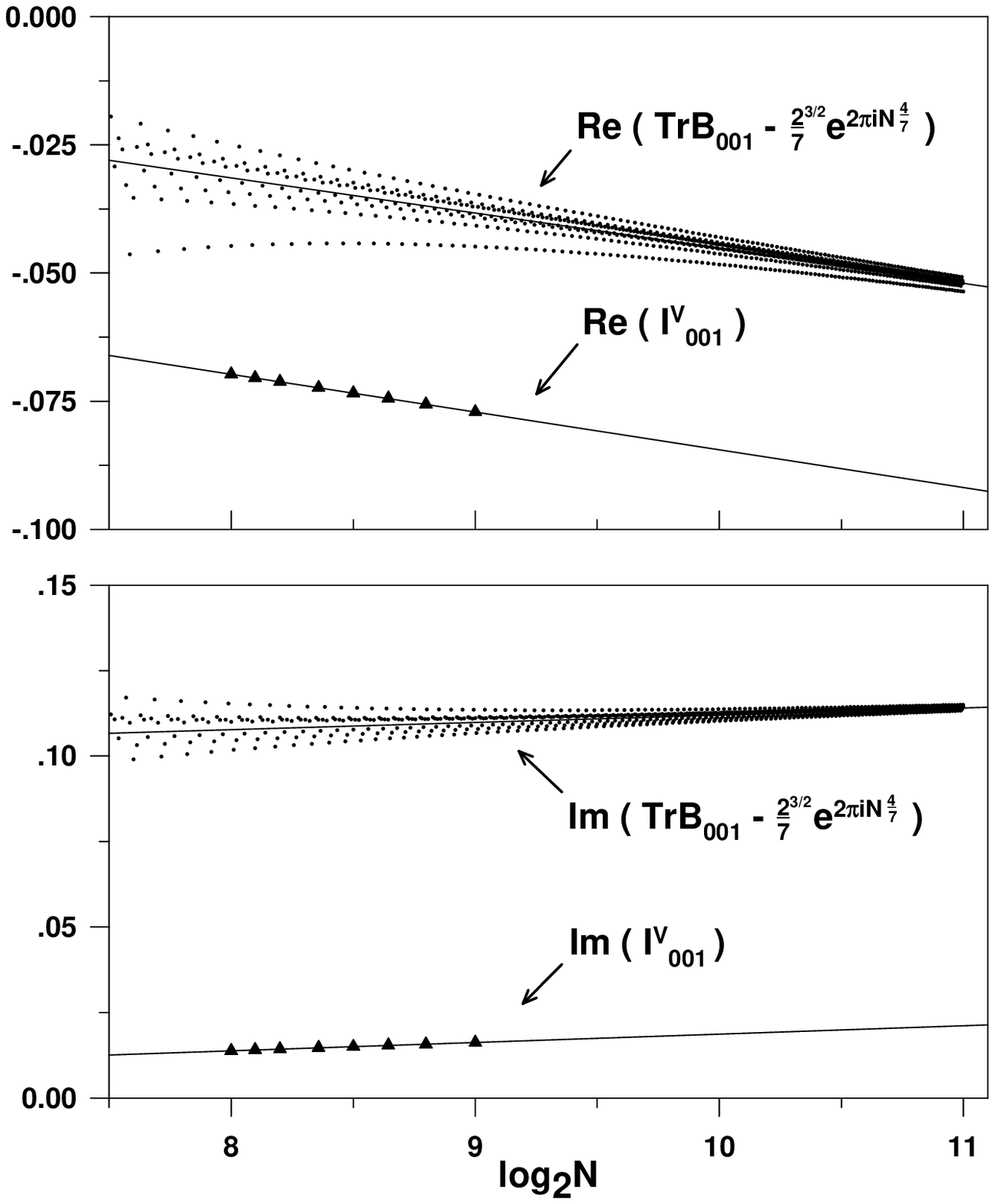}
\caption{ The exact trace $\mbox{Tr}B_{(0,0,1)}$ as a
function of $\log_2 N$ (the Gutzwiller term has been substracted)
together with the corresponding vertex integral $I_{(0,0,1)}^V$.  }
\end{figure}

\newpage

\begin{figure}
\epsfysize=15.0cm
\epsfbox[80 170 532 632]{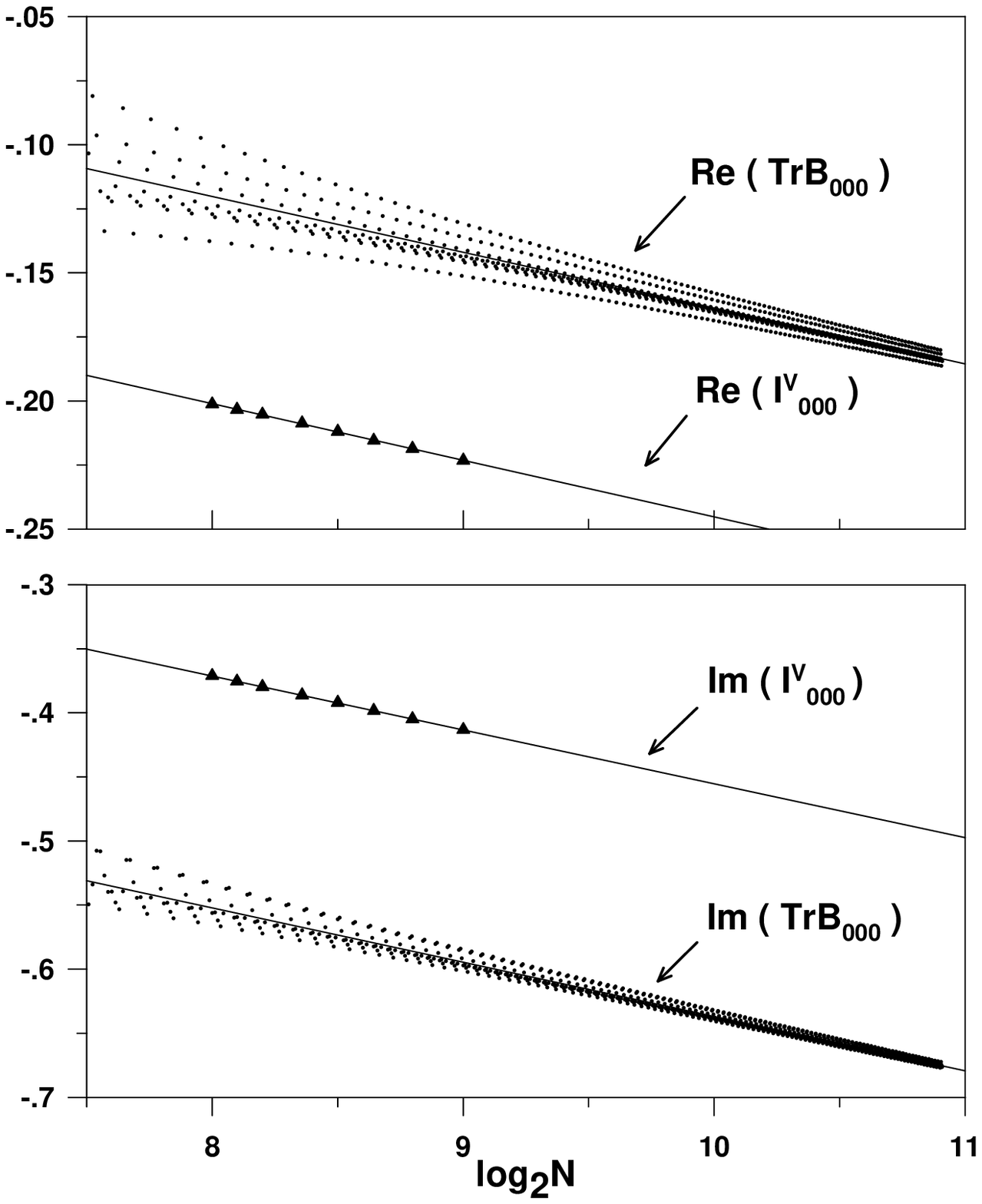}
\caption{ The exact trace $\mbox{Tr}B_{(0,0,0)}$ as a
function of $\log_2 N$ together with the corresponding vertex integral $I_{(0,0,0)}^V$.  }
\end{figure}

\newpage

\begin{figure} 
\epsfxsize=15.0cm
\epsfbox[80 170 542 622]{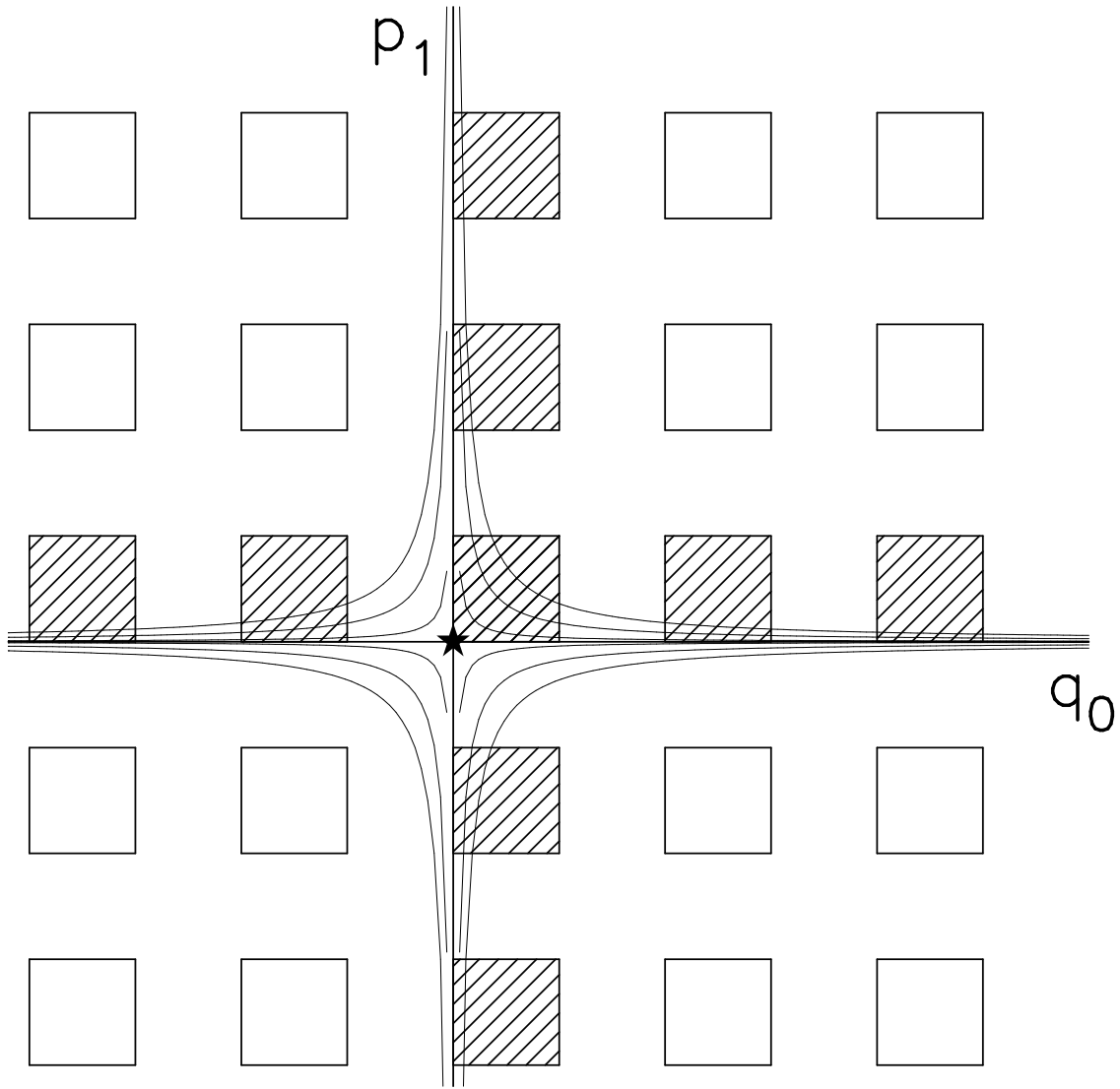}
\caption{ Integration domains. Non--decreasing
contributions to the first trace of the map arise from integrating the
phase $e^{-2 \pi iN p_1 q_0}$ over the shaded squares. The square
containing the origin (a stationary point of the {\em first kind\/})
provides the $\log N$ term plus a constant one. The correct constant
term is obtained when the contributions from all the squares lying
along the axes (they contain an infinity of {\em second kind}
stationary points) are added up.  Several lines of constant phase have
been drawn for reference. } 
\end{figure}

\newpage

\begin{table}[ht]
\caption{Anomalous contributions to the trace from each family of 
paths with a common symbol $\beps$. Displayed are the complex
coefficients $a_{\sbeps}$ of the $\log N$ terms in the 
asymptotic expansion of the three shortest traces.
Exact results were obtained by explicit numerical calculation of the
partial traces. "Theory" means the result of
evaluating the vertex Poisson integral. For times 1--2 the
integrals are done analytically, for $L=3$ numerically. 
In the numerical cases, a linear fit has been 
done to extract the coefficients (See e.g. Figs.~2,~3). The square 
brackets contain the estimated maximum 
error in the least significative figure.}
\vspace{2pc}
\begin{center}
\begin{tabular}{|c|c|c|c|c|c|}                               \hline
 $L$ & $\beps$   & $a_{\sbeps}$ (exact) 
                 & $a_{\sbeps}$ (theory)                \\ \hline
  1  & $(0)$     & $( 0,-0.2251[1])$  
                 & $-i\frac{\sqrt{2}}{2\pi} 
                    \approx ( 0,-0.22508) $              \\ \hline
  2  & $(0,0)$   & $(-0.0234[2] ,-0.106[1] )$  
                 & $-\frac{\log 2 }{3\pi^{2}}-\frac{i}{3\pi} 
                    \approx  (-0.02341,-0.1061 )$ \\ \hline
  2  & $(1,0)$   & $(-0.0232[2] , 0.0001[2] )$  
                 & $- \frac{\log 2 }{3\pi^{2}} 
                    \approx (-0.02341, 0       )$            \\ \hline
  3  & $(0,0,0)$ & $(-0.0313[3] ,-0.0609[5] )$  
                 & $(-0.0318[3] ,-0.0607[2] )$            \\ \hline
  3  & $(0,0,1)$ & $(-0.01055[10] , 0.00350[5] )$  
                 & $(-0.01064[1]  , 0.00352[3])$            \\ \hline
\end{tabular}
\end{center}
\label{tabla}
\end{table}
\end{document}